# Tweets vs. Mendeley readers: How do these two social media metrics differ?

Stefanie Haustein[1,2], Vincent Larivière[1,3], Mike Thelwall[4], Didier Amyot[2]
& Isabella Peters[5]

[1] École de bibliothéconomie et des sciences de l'information, Université de Montréal
C.P. 6128, Succ. Centre-Ville, Montréal, QC. H3C 3J7 (Canada)
*stefanie.haustein@umontreal.ca*

[2] Science-Metrix Inc., 1335 Avenue du Mont-Royal E, Montréal, Québec H2J 1Y6, (Canada)

[3] Observatoire des sciences et des technologies (OST), Centre interuniversitaire de recherche sur la science et la technologie (CIRST), Université du Québec à Montréal
CP 8888, Succ. Centre-Ville, Montréal, QC. H3C 3P8 (Canada)

[4] School of Technology, University of Wolverhampton, Wulfruna Street, Wolverhampton WV1 1LY (UK)

[5] ZBW – German National Library of Economics, Leibniz Information Centre for Economics, Düsternbrooker Weg 120, 24105 Kiel (Germany)

**Abstract**

A set of 1.4 million biomedical papers was analyzed with regards to how often articles are mentioned on Twitter or saved by users on Mendeley. While Twitter is a microblogging platform used by a general audience to distribute information, Mendeley is a reference manager targeted at an academic user group to organize scholarly literature. Both platforms are used as sources for so-called "altmetrics" to measure a new kind of research impact. This analysis shows in how far they differ and compare to traditional citation impact metrics based on a large set of PubMed papers.

**Introduction**

Citations have long been the basis of quantitative research evaluation on the basis that articles that contribute to science will be cited by new research that builds upon them [14]. The social web has introduced new opportunities reflecting impact on a potentially broader audience than publishing authors. Moreover, social media impact can be measured faster than citations as it can occur right after online publication. Many social media sites have Application Programming Interfaces (APIs) that make it easy to access large scale data [19]. This has led to the emergence of a new research area, altmetrics, which focuses on the creation, evaluation and use of scholarly metrics derived from the social web [19]. Uptake from publishers and surveys among researchers suggest that altmetrics are set to become a standard part of the scholarly landscape [1; 6].

Even though it has become apparent that they reflect different things [12], little is known about how various social media metrics differ and what kind of impact they reflect. At the same time, the need to define and classify, to put them in a suitable indicator space ("iSpace", [4]) is increasing. Similar to the development of the Science Citation Index by Eugene Garfield in the 1960s which enabled large-scale citation analysis, both qualitative sociological and quantitative bibliometric research is needed to understand the meaning of these new metrics. Estimates by Adie and Roe [1] suggest that social media activity around scholarly articles is growing by 5% to 10% per month. Focusing on Mendeley and Twitter, this study compares two of the largest social media platforms used for altmetrics and juxtaposes two sources with different purposes and levels of engagement. The comparison is carried out from a quantitative point of view, analyzing the number of citations as well as users who saved journal articles to Mendeley or mentioned them in tweets. The quantitative study is complemented by an explorative analysis of the most frequently read and tweeted documents in two fields of research, providing background information on the topic and demographic information of readers and tweeters.

**Literature Review**

The social referencing site Mendeley [7; 9] as well as similar sites Zotero, CiteULike and Connotea [10; 20] allow users to publically store and share references online. Since these users do not have to be publishing academics, but may also be practitioners or students, these references may be useful as a source of evidence of the wider use of articles. Mendeley currently seems to be the most widely used of these sites and provides a free API enabling the collection of usage data. Even though articles may often be registered in the site without being read, a Mendeley reader count may give an indication of how widely read an article is [2]. These numbers are likely to be significant underestimates since only a small minority of people who read articles are active Mendeley users.

Some studies have used correlation tests to assess whether Mendeley readership figures associate with citation counts for the same articles, with moderate positive results. One study has shown that Mendeley readership figures for Nature and Science articles published in 2007 correlate significantly and moderately with their citation counts [11] and a study of 62,647 articles in five social science disciplines and 14,640 articles published in five humanities disciplines in 2008 found statistically significant and positive correlations in all cases, with low to moderate correlations for each discipline [15]. This gives substantial evidence that Mendeley readership counts are useful measures but reflect a different type of impact than traditional citation counts. Although it has been shown that a large share of Mendeley users consists of PhD students and postdoctoral researchers [16; 21; 26], more research is needed to find out whether it reflects actual readership and by whom.

Unlike Mendeley, Twitter is widely used outside of academia and thus seems to be a particularly promising source of evidence of public interest in science. Hence, tweets to articles likely reflect impact different from both traditional citations and Mendeley readership counts. Nevertheless, Twitter is also used by academics [2; 18; 22; 25] and so tweet counts may reflect educational or scholarly impact as well. One problem with a detailed investigation of tweets of academic articles is that, being restricted to 140 characters, tweets typically provide little context [24] and so it is difficult to be sure why an article has been tweeted.

A small study has given partial evidence of the value of tweet counts as an early indicator of scientific impact for an online open access medical informatics journal by showing that the number of tweets to articles predicted future citations [5]. A later study used the sign test to investigate the relationship between tweets and citations avoiding biases caused by publication age, showing that more tweets are associated with more citations for PubMed articles [23]. However, correlations were very low and even negative for some fields of research [8]. Some of the most tweeted PubMed papers were either topical (e.g., Fukushima), related to scholarly communication in general or about curious topics (e.g., penile fracture) [8]. Further qualitative studies are thus needed to identify the reasons why links to scholarly articles are tweeted (e.g., positive, negative or neutral comments, self-promotion, topicality, curiosity) and by whom (e.g., scientific community, science communicators, general public).

**Methods**

*Data collection*. The comparison of Mendeley reader counts and tweets was based on a set of 1.4 million journal papers (articles and reviews) published between 2010 and 2012, which were both covered in PubMed and Web of Science (WoS) [8]. The WoS documents were classified using the National Science Foundation (NSF) journal classification system, which assigns each paper to one discipline and sub-ordinate specialty. The mapping of PubMed to WoS was carried out on a local copy of WoS matching documents based on bibliographic information. Citations covered those received until calendar week 34 of 2013, so that the citation window varied between 7 to 43 months. Citation counts were not normalized to compare them directly to tweets and reader counts. Mendeley readership data was collected via the Mendeley API and Twitter data were provided by Altmetric [23; 1].

Since unique identifiers were often missing in Mendeley, the retrieval of readership data is based on the title, last name of the first author and publication year. Since the API does not allow for exact match searches, 20 results ranked by relevance to the query have been returned, so that the results had to be checked regarding the number of correct results (if any). As the bibliographic information in the Mendeley database contained many spelling variations and errors in titles and author names that were used for the matching, we allowed similar titles and author names (used in addition if title was shorter than 70 characters and 5 words). This was done by applying a Levenshtein distance which varied depending on the length (5%). These settings proofed to be the best compromise between recall and precision. In a manner similar to tweets, Mendeley data only includes papers with at least one reader. Entries without readers were disregarded. If a paper was represented by more than one entry in Mendeley, the number of readers was aggregated.

*Data analysis*. Mendeley ($P\%_{read}$) and Twitter coverage ($P\%_{tweeted}$), that is the number of papers with at least one Mendeley reader or tweet, and the mean reader ($R/P_{read}$) and Twitter citation rate ($T/P_{tweeted}$) were calculated for all NSF disciplines and specialties based on the 1.4 million documents. Due to the biases caused by citation delay and social-media uptake as described by [23] and [8], Spearman correlations between citations, Mendeley readers and tweets were only calculated for the subset of papers published in 2011.

On the specialty level, data are presented in the framework introduced by [8], which in a coordinate system juxtaposes the breadth of the activity of the field on the social media platforms (coverage above or below average; x-axis) and the similarity between citations and social media counts (positive or negative Spearman correlations; y-axis) identifying four cases: large share of papers appear on platform, usage resembles citing patterns (I), small share of papers on platform, usage resembles citing patterns (II), large share of papers on platform, usage differs from citing patterns (III) and small share of papers on platform, usage differs from citing patterns (IV). The size of data points indicates the intensity with which papers are used, that is the mean number of social media counts for papers with at least one social media count.

Differences between citations, Mendeley readers and tweets were analyzed in more detail for the two specialties General Biomedical Research and Public Health. The three most tweeted and read papers were identified and described to explore the most popular topics on the two platforms. Information on demographics of Twitter and Mendeley users were used as background information to find out by whom and where tweets and reader counts were generated. The three most frequent Mendeley readership categories regarding academic status, country and discipline were available through the API [7]. The Twitter demographics were obtained from Altmetric and indicated the countries of tweeters and, based on keywords in profile descriptions, classified them as members of the public, scientists, medical practitioners and science communicators [13; http://www.altmetric.com/sources-twitter.php].

*Limitations*. Automated data collection is not perfect. The tweeters were retrieved by Altmetric between July 2011 and December 2012 and were limited to tweets published during that time containing a unique identifier (e.g., PMID, DOI, URL). Informal mentions of papers were not captured. Especially for papers published before 2011, Twitter data may be incomplete. Due to these limitations, we assume that the coverage of PubMed papers on Twitter is underestimated [8]. Since correlations and average Twitter citation rates were only computed for papers with at least one tweet, these results should be less affected.

Similarly the collected coverage of Mendeley readership might be underestimated due to server timeouts or the suppression of documents with one reader by the API [7]. The retrieval of Mendeley data is based on matching similar titles and author names using the Levenshtein distance to allow for different spelling variants and typos. Increasing recall is accompanied by reducing precision so that our data set included a small number of false positives. The matching algorithm was tested on a random sample of 401 papers by checking the result lists manually, showing that the automatic matching resulted in 1.7% false positives and 0.7% false negatives. Note that Mendeley improved the quality of data entries in October 2013 merging different entries of the same document, so that a less complex search algorithm might be sufficient in the future. Given that we aggregated readership counts of multiple entries, our results should be comparable to the improved Mendeley database.

**Results and discussion**

The share of PubMed papers with at least one Mendeley reader (66.2%, Table 1) was much higher than the coverage on Twitter (9.4%), although the latter increased significantly over the three years (2.4%[1]; 10.9%; 20.4%), while Mendeley coverage went slightly down (70.3%; 66.5%; 57.3%). The same applies to the intensity with which papers were (re)used once they were on the platform. The sum of Mendeley readers was 9.2 million and the mean reader rate (9.7) was almost four times as high as the mean Twitter citation rate (2.5). Similar to the citation delay, the mean reader rate decreased from 2010 to 2012 papers, although not as significantly (10.7; 9.5; 7.6). It was also always higher than the mean and median citation rate.

While the 390,190 papers from 2011 with at least one Mendeley reader correlated moderately with citations ($\rho=0.456$**, Table 1), the correlation between tweets and citations for 63,800 tweeted papers was much lower ($\rho=0.157$**). Compared to [8], which was based on a seven month shorter citation window, the latter value even decreased. Mendeley and Twitter showed slightly higher but still low correlations ($\rho=0.275$** for 45,229 papers with R>0 and T>0). This confirms assumptions and previous findings [2; 15; 26] that Mendeley measures impact similar but not identical to citations and also shows that tweets and citations are only very weakly associated.

---

[1] This might be partially explained by the underestimation of younger papers described in the limitations section.

**Table 1.** Number of papers in PubMed, share of papers with at least one Mendeley reader ($P\%_{read}$) or one tweet ($P\%_{tweeted}$), sum of Mendeley readers ($R$) or tweets ($T$) and mean number of readers ($R/P_{read}$) or tweets ($T/P_{tweeted}$) per paper with at least one reader or tweet.

| NSF disciplines | $P_{PubMed}$ | $P\%_{read}$ | $P\%_{tweeted}$ | $R$ | $T$ | $R/P_{read}$ | $T/P_{tweeted}$ | $\rho\ P_{read}$ | $\rho\ P_{tweeted}$ |
|---|---|---|---|---|---|---|---|---|---|
| Arts | 71 | 66.2% | -- | 128 | -- | 2.7 | -- | -0.209 | -0.645 |
| Biology | 61,785 | 72.7% | 7.1% | 570,713 | 9,634 | 12.7 | 2.2 | 0.448** | 0.142** |
| Biomedical Research | 286,398 | 72.4% | 9.8% | 2,973,664 | 90,633 | 14.3 | 3.3 | 0.530** | 0.232** |
| Chemistry | 121,874 | 60.8% | 5.5% | 619,418 | 10,933 | 8.4 | 1.6 | 0.476** | 0.147** |
| Clinical Medicine | 779,707 | 62.8% | 10.1% | 3,712,112 | 184,002 | 7.6 | 2.4 | 0.439** | 0.155** |
| Earth and Space | 26,938 | 72.4% | 4.0% | 155,095 | 2,885 | 8.0 | 2.7 | 0.396** | 0.082 |
| Engineering and Technology | 27,792 | 71.6% | 5.5% | 304,512 | 2,916 | 15.3 | 1.9 | 0.622** | 0.159** |
| Health | 59,073 | 67.0% | 12.8% | 257,973 | 17,306 | 6.5 | 2.3 | 0.336** | 0.099** |
| Humanities | 691 | 40.7% | 6.5% | 1,036 | 121 | 3.7 | 2.7 | 0.227** | 0.007 |
| Mathematics | 2,461 | 69.2% | 5.4% | 13,586 | 197 | 8.0 | 1.5 | 0.306** | -0.209 |
| Physics | 19,892 | 76.4% | 1.8% | 124,904 | 539 | 8.2 | 1.6 | 0.386** | 0.032 |
| Professional Fields | 5,600 | 72.1% | 17.0% | 45,231 | 2,510 | 11.2 | 2.6 | 0.370** | 0.177** |
| Psychology | 35,980 | 81.0% | 14.9% | 408,440 | 16,240 | 14.0 | 3.0 | 0.441** | 0.075** |
| Social Sciences | 9,019 | 68.8% | 9.1% | 54,253 | 2,192 | 8.7 | 2.7 | 0.431** | 0.054 |
| **Total** | **1,437,281** | **66.2%** | **9.4%** | **9,241,065** | **340,751** | **9.7** | **2.5** | **0.456** | **0.157** |

Table 1 indicates the Mendeley and Twitter coverage, mean reader and Twitter citation rate and correlations with citations on the level of NSF disciplines. As these results are based on PubMed, they only represent papers with a biomedical focus. That is, results for Engineering and Technology are not representative for the whole discipline but those papers that are covered by PubMed. Keeping this bias in mind, Psychology had the highest coverage on Mendeley and the second highest on Twitter. Humanities papers appeared least frequently on Mendeley and were also covered below average on Twitter, although Physics provided the papers with the lowest coverage on Twitter. The intensity of (re)use was highest for Engineering and Technology, Biomedical Research and Psychology on Mendeley and Biomedical Research, Psychology, Social Science, Humanities and Earth and Space on Twitter. Correlations between citations and Mendeley readers were significant for all disciplines except for Arts ($\rho=-0.209$), but varied between 0.227** (Humanities) and 0.622** for Engineering and Technology. A different picture presents itself analyzing correlations between the number of citations and tweets. No significant correlations were found for six disciplines, of which Arts ($\rho=-0.645$) and Mathematics ($\rho=-0.209$) were negative. Significant positive correlations were also below those of Mendeley ranging from 0.075** (Psychology) to 0.232** (Biomedical Research).

On the level of research specialties, differences between Mendeley and Twitter become even more apparent. Using the same scales for the frameworks (Figure 1) representing Mendeley (A) and Twitter (B) activities, it can be seen that differences between specialties are much less pronounced on Mendeley than on Twitter. All specialties showed positive correlations with citations, so that all 119 specialties were classified as either Case I or Case II, meaning that Mendeley reading patterns resembled citation patterns but coverage was either above (I) or below average (II). Spearman values differed between specialties, ranging from History ($\rho=0.038$) and Psychoanalysis ($\rho=0.136$) to Social Psychology ($\rho=0.621$**), Embryology ($\rho=0.625$**), Management ($\rho=0.631$**), General Biomedical Research ($\rho=0.677$**) and Materials Science ($\rho=0.682$**). Psychoanalysis and History were also the fields with the lowest coverage. As 46 other specialties they were covered below the overall PubMed average of 66.2%, while 71 were above. The highest shares of papers were read in Experimental Psychology ($P\%_{read}=85.6\%$), Geology (88.0%) and Ecology (88.3%). The highest reader rate, which is indicated by the size of data points, was obtained by papers in Ecology, General Biomedical Research and Management, which shows that coverage and activity on Mendeley seem to influence each other positively (Pearson's r=0.566). As shown by the overall mean Twitter citation rate in Table 1, tweet activity per field was also much lower than readership activity on Mendeley. It ranged from 1.1 (Polymers) to 6.5 (General Biomedical Research). General & Internal Medicine (4.5), Social Psychology (4.4) and Experimental Psychology (4.3) follow as the second to fourth most active fields on Twitter. Compared to Mendeley, Twitter coverage was not only much lower but also much more skewed compared to the average of 9.4%. The highest share of papers is tweeted in Communication (27.6%), Social Psychology (25.5%), Anesthesiology (21.5%), Management (21.0%) and Nutrition & Dietic (20.4%). Topics related to these fields could be relevant to a broader audience or general public. Correlations between citations and tweets are much lower than those with reader counts and even negative for 14 specialties, i.e. Pharmacy (Case IV: $\rho=-0.173*$; $P\%_{tweeted}=4.5\%$) and Speech-Language Pathology and Audiology (Case III: $-0.175$; 9.7%). Spearman values are highest for Anthropology and Archeology (Case II: 0.372**; 8.7%) and Genetics and Heredity (Case I: 0.279**; 11.2%).

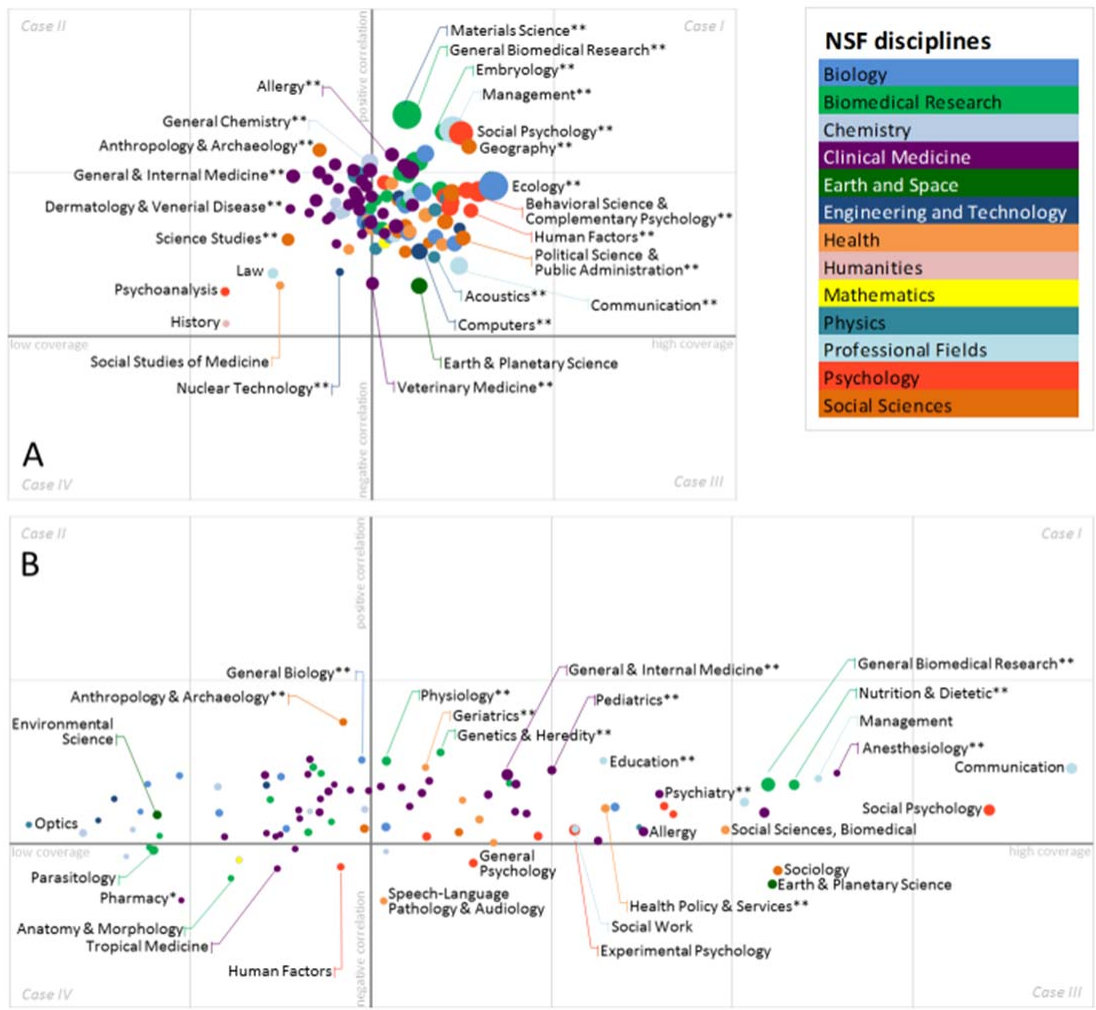

**Figure 1:** Frameworks representing Mendeley (A) and Twitter data (B) on the level of NSF specialties. Specialties are shown if they were represented in PubMed by more than 100 papers for the 2010-2012 period and more than 30 read (A) or tweeted (B) papers in 2011.

The two specialties General Biomedical Research (NSF discipline of Biomedical Research) and Public Health (Health) were selected for an exploratory study because they represent extreme cases.[2] Given their positions in the Twitter and Mendeley frameworks (Fig. 1), one would expect a broad interest by the general public for Public Health, because a large share of papers was mentioned on Twitter but there was no correlation between tweeting and citing patterns [8], and a large interest by an academic audience in General Biomedical Research, since Mendeley coverage was above average, reader counts correlated strongly with citations and the reader rate per document was 3.6 times higher than average. Twitter citation rate, coverage and correlations with citations of General Biomedical Research were also above average, while Public Health papers were saved less frequently on Mendeley than average and correlations with citations were also below the average of 0.456.

To further explore by whom the most popular papers of the two specialties were saved on Mendeley and mentioned on Twitter, the three most read and tweeted papers are described in more depth. The paper with the most tweets in General Biomedical Research[3] analyzed the effects of the Chernobyl accident on thyroid cancer and was published online in May 2011, two month after the nuclear catastrophe in Fukushima. As indicated by the Twitter demographics information by Altmetric, the majority of the 963 people who (re)tweeted this paper were members of the general public (over 90%) in Japan (almost 60%). Similar demographics apply to the paper with the second highest number of tweeters (639)[4], which describes the radioactive soil contamination in Japan after Fukushima. In these two cases, the high popularity on Twitter in combination with the demographic information suggests that a general public discusses scientific papers related or relevant to a topical issue, i.e. a nuclear catastrophe. The third article on the effects of constant access to information on memory (558 tweeters)[5] was also discussed by the general public although less pronounced (approx. 75%) and the readership is more international with one quarter from the US.

The three papers with the most Mendeley readers from the field of General Biomedical Research had 1,153 readers[6], 956 readers[7] and 903 readers[8]. The first appeared in Science introducing a new statistical method for the analysis of large data sets. The majority of its readers were young researchers (37% PhD students, 17% Postdocs) and came from the US (39%), UK (7%) and China (6%). In terms of disciplines, the readership varied, with the three most frequent (Biological Sciences, 46%; Computer and Information Science, 16%; Physics, 5%) making up 77%, which might indicate an interdisciplinary interest in this method. The absence of Master's students, which are often among the top three academic statuses on Mendeley, might indicate that this method is too new to be adapted in teaching. The paper has been cited 21 times on WoS and also been intensively discussed on Twitter (158 tweets;

---

[2] Due to space limitations figures of scatterplots depicting the number of citations, Mendeley readers and tweeters could not be included but can be found online (http://dx.doi.org/10.6084/m9.figshare.1007654) and in the Appendix of this preprint.

[3] Hess, J. et al. (2011). Gain of chromosome band 7q11 in papillary thyroid carcinomas of young patients is associated with exposure to low-dose irradiation. *PNAS*, 108(23), doi: 10.1073/pnas.1017137108.

[4] Yasunari, T.J. (2011). Cesium-137 deposition and contamination of Japanese soils due to the Fukushima nuclear accident. *PNAS*, 108(49), doi: 10.1073/pnas.1112058108.

[5] Sparrow, B., Liu, J., & Wegner, D.M. (2011). Google Effects on Memory: Cognitive Consequences of Having Information at Our Fingertips. *Science*, 333(6043), 776-778, doi: 10.1126/science.1207745.

[6] Reshef, D.N. et al. (2011). Detecting novel associations in large data sets. *Science*, 334(6062), 1518-1524, doi: 10.1126/science.1205438.

[7] Grabherr et al. (2011). Full-length transcriptome assembly from RNA-Seq data without a reference genome. *Nature Biotechnology*, 29(7), 644-652, doi: 10.1038/nbt.1883.

[8] Arumugam et al. (2011). Enterotypes of the human gut microbiome. *Nature*, 473(7346), 174-180, doi: 10.1038/nature09944.

almost 50% members of the public, more than 40% scientists). This paper has thus had impact on both the scientific community and a broader audience on Twitter. The second publication appeared in Nature Biotechnology introduces a new method in genome sequencing and, unsurprisingly, the great majority of its Mendeley readers where from the Biological Sciences (89%), reflecting a high level of specialization. One third of Mendeley readers were PhD students, 17% Postdocs and 9% Master's students and mainly came from the US (32%), UK (7%) and Germany (7%). The paper was also highly cited (107) and has been mentioned by 16 tweets, 82% of which were identified by the Altmetric demographic algorithm as scientists. Its impact can thus be described as mainly scientific. The third paper with 903 Mendeley readers identified three types of people according to the bacterial compositions in their guts, a kind of classification that can be compared to blood types. Readers on Mendeley were mostly PhD students (28%), Postdocs (15%) and researchers (9%), came from the US (36%), UK and France (both 7%) and had a biological (75%), medical (12%) or computer and information science (3%) background. With 292 citations, the paper also had a large impact on the scientific community and with 25 tweets has also been mentioned on Twitter by members of the public (62%), scientists (24%), medical practitioners (10%) and science communicators (3%) from the UK, Japan (both 10%) and the US (6%).

For the NSF specialty of Public Health, the most tweeted paper[9] presents a meta-analysis of studies analyzing the relation between social factors and adult mortality, which identified low education, racial segregation, low social support and poverty as factors that increase mortality risks. About half of the 78 tweeters were from the US, which was to be expected given the national focus of the paper. The majority of Twitter users linking to the paper were members of the public (58%), followed by medical practitioners and health professionals (17%) and scientists (15%), while 8% of tweets were sent by science journalists, bloggers or editors. The paper has also been cited 12 times and had 37 readers on Mendeley. A paper[10] on the reactions to a press release that described the use of pig's blood in cigarette filters was tweeted almost completely by members of the public and almost half of them were from Indonesia. This can be explained by a particularly high interest in the topic by Muslim smokers. Indonesia is the country with the largest share of the world's Muslim population [17]. This paper had hardly any impact on the scientific community. The third most tweeted publication from Public Health[11] showed that the use of non-branded cigarette packs could reduce consumption and increase avoidant behavior by young adult smokers. Of the 49 tweeters, two-thirds were members of the public, 14% medical practitioners, 10% scientist, 8% science communicators and 36% came from Australia, 12% from the UK and 6% each from New Zealand and the US. The study received 6 citations and was not yet saved on Mendeley.

The three Public Health papers with the highest reader counts have 142 readers[12], 92 readers[13] and 90 readers[14]. The first estimated the number of occurrences of illness, hospitalization and death caused by Salmonella, norovirus and other foodborne illnesses. Of its 142 Mendeley readers, 29% were Master's and 24% PhD students mainly from the US (52%) and Biological Sciences (67%; Medicine, 17%; Engineering, 7%). At 377 citations since its publication in 2011, this paper was heavily cited. Tweets could not be captured by Altmetric, as the journal does not use DOIs. The second paper is a Cochrane review addressing the effectiveness of lay health workers' in the areas of maternal and child health and infectious diseases. Readership data was more evenly distributed across countries with 27% of the 92 readers coming from the US, 19% from the UK and 9% from South Africa. One third were Master's students, 13% researchers and 12% PhD students. More than half of them had a medical (54%) and almost one third a Social Sciences (29%) background. The paper has not yet been cited or tweeted. Basch's publication in the Journal of School Health claims that "Healthier students are better learners" and was saved by 90 Mendeley users, 27% of which were Master's and 23% PhD students and 9% were classified as other professionals. Both in terms of countries and disciplines, readers were less skewed but more evenly distributed than for examples shown above with 42% and 40% making up the three most frequent categories, respectively. These top three were US (26%), UK (9%) and South Africa (7%) for countries and Social Sciences (16%), Environmental Sciences (13%) and Business Administration (11%) for discplines. Given the title in combination with a high share of Master's students and distribution of readers among many disciplines (top 3 make up 40%), suggests that this paper might have been read more out of a personal than research interest.

---

[9] Galea, S., Tracy, M., Hoggatt, K.J., Dimaggio, C., & Karpati, A. (2011). Estimated deaths attributable to social factors in the United States. *American Journal of Public Health*, 101(8), 1456-1465, doi: 10.2105/AJPH.2010.300086.

[10] MacKenzie, R., & Chapman, S. (2011). Pig's blood in cigarette filters: how a single news release highlighted tobacco industry concealment of cigarette ingredients. *Tobacco Control*, 20(2), 169-172, doi: 10.1136/tc.2010.039776.

[11] Moodie, C., Mackinthosh, A.M., Hastings, G., & Ford, A. (2011). Young adult smokers' perception of plain packaging: a pilot naturalistic study. *Tobacco Control*, 20(5), 367-373, doi: 10.1136/tc.2011.042911.

[12] Scallan, E.S. et al. (2011). Foodborne Illness Acquired in the United States. *Emerging Infectious Disesase*, 17(1), 7-15.

**Conclusion and outlook**

The comparison of citations with Mendeley reader counts and tweets provides empirical evidence that these three measures are indicators of different types of impact and on different social groups. While citations reflect the impact of papers on the scientific community (i.e., on scientists who are themselves producers of research papers), Mendeley seems to mirror the use of these papers by a broader but still largely academic audience, which currently consists to a large extent of students and postdocs [16; 26]. Whereas low correlations and frequently tweeted topics suggest that Twitter reflects the popularity among a general public. Both content analyses and qualitative user studies are needed to confirm these assumptions to determine if they hold for all fields of science or, especially with regards to Twitter, if only specific topics are discussed by the general public. More specifically, the large-scale quantitative study has shown that a) the number of Mendeley readers and tweets are two distinct social media metrics and they differ from citations, and b) differences in breadth of distribution, intensity and correlation with citation patterns differ between specialties for both metrics. These two results imply that a) one social media metric is not like the other and by no means are they an alternative to citation impact measures, and b) social media counts of papers from different fields of research are not directly comparable, a fact long known in traditional bibliometrics.

The exploratory descriptions of highly tweeted and read papers suggest that some had more academic and scientific impact whereas others were highly relevant to a more general public, while some had both. It was also shown that tweets often seem to represent discussions by "members of the public", at least as defined by Altmetric's demographic algorithm, but this was not always the case. This emphasizes that the Twittersphere also contains scientists, teachers, students, librarians, funders, politicians etc. discussing scholarly contents, who may or may not act according to these roles when they tweet. As Lin and Fenner [12] wrote, it is not yet possible to differentiate between scholarly and non-scholarly impact based merely on the platform used. Moreover, the level of engagement does not only differ between (saving a paper to a library vs. tweeting about it) but also within platforms (saving vs. reading and retweeting a link vs. discussing the content).

---

[13] Lipp, A. (2011). Lay Health Workers in Primary and Community Health Care for Maternal and Child Health and the Management of Infectious Diseases: A Review Synopsis. *Public Health Nursing*, 28(3), 243-245, doi: 10.1111/j.1525-1446.2011.00950.x.

[14] Basch, C.E. (2011). Executive summary: Healthier students are better learners. *Journal of School Health*, 81(10), 591-592, doi: 10.1111/j.1746-1561.2011.00631.x.

The differentiation between audiences and engagements needs to be subject of future research. A first approach on automatic identification of scientists on different social media sites was proposed by [3] and needs to be further investigated. Large-scale quantitative analyses have to be combined with content and context analyses as well as qualitative research to investigate by whom and how social media counts are generated. This will reveal the reasons for disciplinary differences (varying interests in topics vs. social media affinity of scientific communities) and help determine what it is that social media metrics actually measure: social impact, scientific impact or buzz.

## Appendix

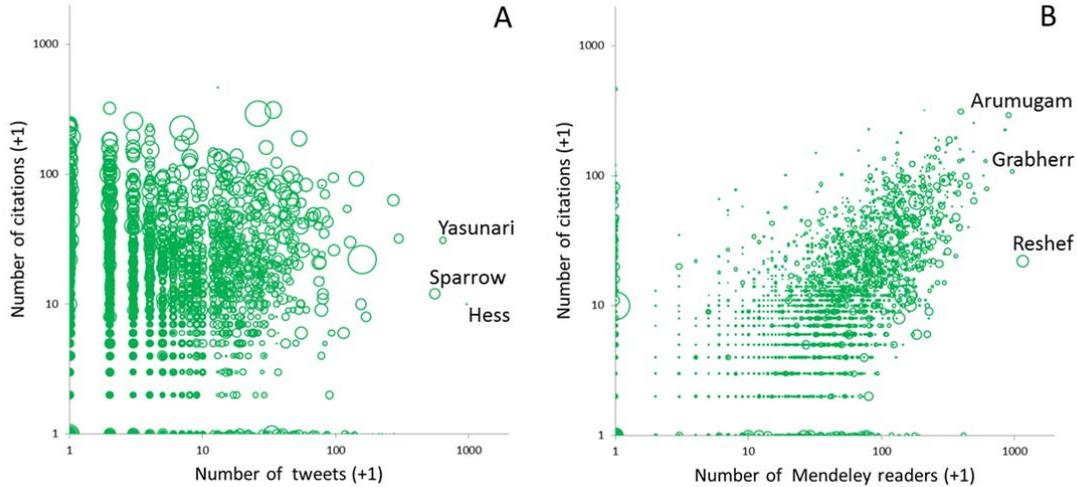

**Figure A1:** Scatterplot of number of citations and number of tweets (A, $\rho=0.181**$) and Mendeley readers (B, $\rho=0.677**$) for papers published in General Biomedical Research in 2011. The respective three most tweeted (A) and read (B) papers are labeled showing the first author. All values were increased by 1 to include all data points in the logarithmic representation.

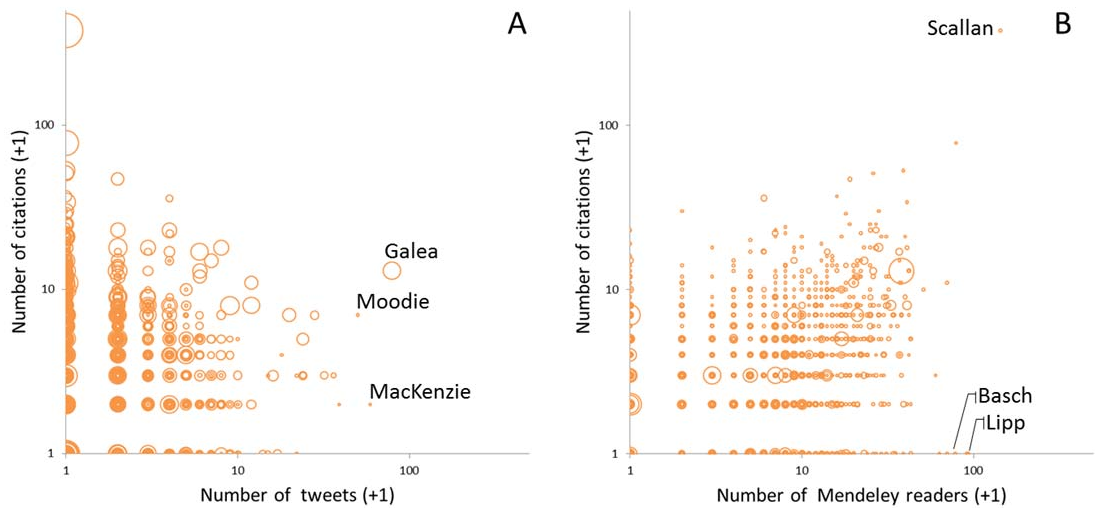

**Figure A2:** Scatterplot of number of citations and number of tweets (A, $\rho=0.074**$) and Mendeley readers (B, $\rho=0.351**$) for papers published in Public Health in 2011. The respective three most tweeted (A) and read (B) papers are labeled showing the first author. All values were increased by 1 to include all data points in the logarithmic representation.

*Figures can also be found online: http://dx.doi.org/10.6084/m9.figshare.1007654.*